\documentclass[%
 preprint, 
 amsmath,amssymb,
 aps, physrev,
]{revtex4-2}

\usepackage{graphicx}
\usepackage{dcolumn}
\usepackage{bm}
\usepackage{xcolor}

\begin{document}

\preprint{APS/123-QED}

\title{\textbf{Attenuation of long-wavelength sound in quenched disordered media} 
}%

\author{Bingyu Cui}
\altaffiliation{School of Science and Engineering, The Chinese University of Hong Kong (Shenzhen), Longgang, Shenzhen, Guangdong, 518172, P.R. China.}
 \email{bycui@cuhk.edu.cn}
\author{Yuqi Wang}
\affiliation{School of Science and Engineering, The Chinese University of Hong Kong (Shenzhen), Longgang, Shenzhen, Guangdong, 518172, P.R. China.}%

\date{\today}

\begin{abstract}
We derive analytically, and validate numerically, the dispersion renormalization and attenuation of acoustic waves propagating through quenched disordered media in the long-wavelength limit. We consider weak spatial fluctuations in elastic moduli and/or mass density and compute the disorder-induced self-energies within the leading (Born) approximation. For sufficiently weak disorder, the results depend only on the variances of the fluctuations and are therefore insensitive to the detailed form of the underlying random distribution. For spatially uncorrelated elasticity disorder we obtain Rayleigh-type attenuation, $\Gamma(q)\propto q^{d+1}$ , together with a reduction of the sound speed. In contrast, density disorder produces Rayleigh-type attenuation but does not renormalize the acoustic dispersion to leading order. Molecular dynamics simulations and normal-mode analyses of disordered one- and two-dimensional lattices quantitatively confirm the theoretical predictions.
\end{abstract}

\maketitle


\section{\label{sec:level1}Introduction}

The propagation of sound waves in perfect crystals at low temperature is well understood, while the underlying mechanism of sound attenuation in disordered solids remains debated. Longitudinal and transverse sound attenuation can be measured in light or neutron scattering experiments \cite{andersonAnomalousLowtemperatureThermal1972a,benassiEvidenceHighFrequency1996,hehlenHyperRamanScatteringObservation2000,buchenauLowfrequencyModesVitreous1986,buchenauNeutronScatteringStudy1984,ruffleObservationOnsetStrong2003,foretScatteringInvestigationAcoustic1996,malinovskyNatureBosonPeak1986,zellerThermalConductivitySpecific1971,Sette1998,Masciovecchio2006,Delaire2011,Li2020}. At small wavenumber $q$, attenuation coefficients exhibit Rayleigh-type scattering, $\Gamma(q)\sim q^{d+1} (d$ the spatial dimension) due to solid's inhomogeneities \cite{zellerThermalConductivitySpecific1971,Zaitlin1975,Mizuno2018,Wang2019,Moriel2019}, although sometimes a logarithmic correction to this scaling is reported \cite{Gelin2016,Mizuno2018,Cui2020,Ding2025}.

Proposed models for the physical mechanism of sound attenuation in low-temperature disordered solids include ``isotopic scattering'' model \cite{zellerThermalConductivitySpecific1971}, ``local oscillator'' \cite{Buchenau1992,Schirmacher2011}, random matrix models \cite{Grigera2011,Conyuh2020,Ganter2010} and nonaffine effects \cite{Szamel2022}, which are used to derive the Rayleigh scattering. Simulations and local elastic measurements imply that strongly heterogeneous moduli, nonaffine displacements, and mesoscopic elastic correlations are key ingredients in sound attenuation phenomenology~\cite{Schirmacher2010,Schirmacher2008,sokolovMediumrangeOrderGlasses1992,mizunoAcousticExcitationsElastic2014,Gelin2016,Cui2020,Baggioli2022,Lindsay2023,Szamel2022,Szamel2025}. Motivated by this evidence, heterogeneous-elasticity theory (HET) based on the replica trick~\cite{schirmacherAcousticAttenuationGlasses2007,schirmacherHarmonicVibrationalExcitations1998,marruzzoHeterogeneousShearElasticity2013,schirmacherHeterogeneousViscoelasticityCombined2015,schirmacherThermalConductivityGlassy2006a}, related Euclidean-random-matrix  (ERM) \cite{Mzard1999,Schirmacher2019,Szamel2025} and coherent-potential-approximation \cite{Schirmacher1998,Economou2006} approaches treat glasses as a random elastic continuum and capture key vibrational anomalies such as Rayleigh-like attenuation~\cite{conyuhApplicationRandomMatrix2019,parisiOriginBosonPeak2003,beltukovRandomMatrixTheory2015,cilibertiBrillouinBosonPeaks2003,mezardSpectraEuclideanRandom1999}, although they do not clarify how far elastic disorder alone can modify long-wavelength acoustic dispersion and attenuation.

Existing HET implementations based on a specific distribution of random elasticity are mostly scalar or purely shear~\cite{marruzzoHeterogeneousShearElasticity2013}; fully vectorial elasticity with distinct longitudinal and transverse branches is only treated approximately, and spatial correlations of the elastic moduli are usually neglected or handled only in specific numerical cases. Fluctuations in densities are also completely ignored in HET. Further, there is no quantitative analysis of the strength of disorder, making it difficult to separate the genuine effects of elastic heterogeneity from Van Hove singularities in crystal physics~\cite{chumakovEquivalenceBosonPeak2011,Chumakov2014}.

In this work we develop a complementary, explicitly perturbative description of long-wavelength acoustic propagation in quenched random media. Instead of relying on replica-based HET or related effective-medium/ERM-type formulations, we compute the disorder-induced self-energies directly within the leading nonvanishing (Born) approximation. This yields closed analytic expressions for the renormalized longitudinal and transverse dispersions and attenuation rates in arbitrary spatial dimension $d$, with a transparent dependence on disorder strength through the second moments (variances or covariances) of Lam\'e-parameter fluctuations, regardless of the type of disorder. The same framework also allows us to treat density disorder on the same footing and to separate its contribution from that of elastic heterogeneity. In the long-wavelength limit and for spatially uncorrelated disorder, we recover Rayleigh-type attenuation, $\Gamma(q)\propto q^{d+1}$, while clarifying the distinct effects on dispersion: Weak elastic heterogeneity reduces the sound speed at leading order, whereas density disorder produces attenuation but does not renormalize the acoustic dispersion to the same order in $q$. We test these predictions using molecular dynamics measurements and normal-mode calculations of the dynamical structure factor in disordered one- and two-dimensional lattices, finding quantitative agreement in the low-$q$ regime and delineating where lattice effects or excitation protocols can generate apparent attenuation.

The remainder of the paper is organized as follows: Section II presents the theoretical derivations. Section III elaborates numerical schemes implemented in this article, followed by explicit validation and discussions in Sec. IV. Section V concludes.

\section{\label{sec:level2} Formalisms}
In this section, we present our approach for the sound wave attenuation in disordered systems.

\subsection{Heterogeneous elastic theory in 1-dimension}
We develop a theory of sound wave propagation in a heterogeneous elastic continuum, beginning with a scalar model in 1-dimensional space. The displacement field $u(x,t)$ obeys
\begin{equation}
  \rho_0\frac{\partial^2 u(x,t)}{\partial t^2} = \frac{\partial}{\partial x}\left[K(x)\,\frac{\partial u(x,t)}{\partial x}\right],
  \label{eq:1d_wave}
\end{equation}
where $\rho_0$ is density, $K(x) = K_0\bigl[1+\delta K(x)\bigr]$ is the elastic modulus with the zero mean fluctuation, $\langle \delta K(x)\rangle=0$ and variance $\langle \delta K(x_0)\,\delta K(x_0+x)\rangle =  f(x)$ for some function $f(x)$, $\langle\cdots\rangle$ denotes an ensemble average. While our analysis applies to arbitrary $f(x)$, in this article, we restrict ourselves to the spatially uncorrelated case, namely $f(x)=\sigma^2\delta(x)2\pi/q_D$, where $q_D$ is the Debye (cutoff) wavenumber such that $\sigma$ is dimensionless and $2\pi$ is a normalization factor.

The fluctuation in elasticity maps to the random scattering potential on the free propagating wave \cite{Economou2006,Altland2010}, where multiple-scattering effects are collected into a self-energy $\Sigma(q,\omega)$ in the ensemble-averaged Green's function $\langle G(q,\omega)\rangle$ in Fourier space, which, by the Dyson equation, is (see Sec. \ref{App:A} in Appendix for detailed derivations)
\begin{equation}
  \left[\langle G(q,\omega)\rangle\right]^{-1} =
  \bigl[G_{0}(q,\omega)\bigr]^{-1} - \Sigma(q,\omega),
  \label{eq:dyson}
\end{equation}
where 
\begin{equation}
  G_{0}(q,\omega) = \frac{1}{-\omega^2 + v_0^2 q^2 + i0^+}
  \label{eq:bareGreen}
\end{equation}
is Green's function for the free propagating wave and $v_0 = \sqrt{K_0/\rho_0}$ is the average speed of sound. To calculate the self-energy, we make the (second-order) Born approximation by keeping only the lowest nonvanishing contribution after averaging, yielding (see details in Sec. \ref{App:A} in Appendix)
\begin{equation}
  \Sigma_{\text{Born}}(q,\omega)
  = \,\frac{\sigma^2}{q_D} \int_0^{q_D}\frac{dp}{2\pi}\,(q\,p)^2\,\,
  G_{0}(p,\omega).
  \label{eq:scalar_sigma_born}
\end{equation}
Note that the Green function and self-energy obtained in this approach are not self-consistent: They are different from the coupled equations obtained in the replica trick formulated upon mapping the random fluctuations in elasticity to a probability distribution \cite{Maurer2004,Schirmacher2008,Cui2020}. Note that the validity of the lowest order Born approximation Eq. \eqref{eq:scalar_sigma_born} is robust to the specific form of the probability distribution in the weak disorder sense, i.e. the fluctuation strength $\sigma^2\ll1$. In the long-wavelength $q\to 0$ limit, upon writing
\begin{equation}
    G_0(p,\omega)=\mathcal{P}\frac{1}{v_0^2p^2-\omega^2}-i\pi\delta\left(v_0^2p^2-\omega^2\right),
\end{equation}
where $\mathcal{P}$ is the principal part, it can be shown that
$\text{Re}\left[\,\Sigma_{\text{Born}}\right] \approx \sigma^2 v_0^2 q^2$ and
$\text{Im}[\,\Sigma_{\text{Born}}] \propto \sigma^2 q^3\,$, leading to the renormalization in the sound dispersion and finite damping. The imaginary part of the Green function becomes
\begin{equation}
    \text{Im}\left[\langle G(q,\omega)\rangle\right]=\frac{\omega\Gamma(q)}{(\omega^2-\Omega^2(q))^2+\omega^2\Gamma^2(q)}
    \label{eq:GreenBorn}
\end{equation}
with
\begin{subequations}
\label{eq:scalar_1d_result}
\begin{align}
  \Omega^2(q)&\equiv v_0^2q^2-\text{Re}[\Sigma_{\text{Born}}]\sim\left(1-\sigma^2\right)v_0^2q^2,\\
  \Gamma(q)&\equiv-\frac{\text{Im}[\,\Sigma_\text{Born}]}{\omega} \propto \sigma^2 q^2.
\end{align}
\end{subequations}
Equation~\eqref{eq:scalar_1d_result} implies that the sound speed is reduced by $\sigma^2/2$, while the damping exhibits a Rayleigh-type scattering together with the disorder variance.

\subsection{Heterogeneous elastic theory in higher dimensions}
For spatial dimension $d\ge 2$, the displacement field is vectorial. The $\alpha$th component of the displacement field, $u_\alpha(\mathbf{r},t)$, obeys
\begin{equation}
  \rho_0\frac{\partial^2 u_\alpha}{\partial t^2} =\frac{\partial}{\partial r_\beta}\left[C_{\alpha\beta\gamma\chi}(\mathbf{r})\frac{\partial u_\chi}{\partial r_\gamma}\right],
  \label{eq:vec_wave}
\end{equation}
where $C_{\alpha\beta\gamma\chi}$ is the elastic tensor. For an isotropic medium, the fourth-rank elastic tensor takes the Lam\'e form \cite{marruzzoHeterogeneousShearElasticity2013,Schirmacher2015t}
\begin{equation}
  C_{\alpha\beta\gamma\chi}(\mathbf{r}) = A(\mathbf{r})\,\delta_{\alpha\beta}\delta_{\gamma\chi}
  + B(\mathbf{r})\left(\delta_{\alpha\gamma}\delta_{\beta\chi}
  + \delta_{\alpha\chi}\delta_{\beta\gamma}\right)
  \label{eq:lame}
\end{equation}
where $A(\mathbf{r})$ is Lam\'e's first parameter and $B(\mathbf{r})$ is the shear modulus. As before, we let the elastic moduli fluctuate around their mean values, $A(\mathbf{r}) = A_0\bigl[1+\delta A(\mathbf{r})\bigr]$ and $B(\mathbf{r}) = B_0\bigl[1+\delta B(\mathbf{r})\bigr]$ with zero means $\langle \delta a(\mathbf{r})\rangle=0$ and $\langle \delta a(\mathbf{r}_0)\delta b(\mathbf{r}_0+\mathbf{r})\rangle=\sigma^2_{ab}\,\,\delta^{(d)}(\mathbf{r})(2\pi/q_D)^d, a,b=A,B$ the correlation between elastic moduli fluctuations. Again, $q_D$ is the Debye wavenumber such that the fluctuation strength $\sigma_{ab}$ are dimensionless. Upon defining the projection operators $P^{L}_{\alpha\beta}(\mathbf{q})= q_\alpha q_\beta/q^2$ and
$P^{T}_{\alpha\beta}(\mathbf{q})= \delta_{\alpha\beta} - q_\alpha q_\beta/q^2$, all tensorial quantities can be decomposed into longitudinal and transverse parts with respect to the wavevector $\mathbf{q}$. Following the same treatment as in the 1-dimensional case, the disorder-averaged Green's tensor takes the form (see detailed derivations Sec. \ref{App:B} in Appendix)
\begin{align}
  \langle G_{\alpha\beta}(\mathbf{q},\omega)\rangle
  &= \frac{P^{L}_{\alpha\beta}(\mathbf{q})}{v_L^2 q^2 - \omega^2 - \Sigma_L(\mathbf{q},\omega)}\nonumber\\
  &+ \frac{P^{T}_{\alpha\beta}(\mathbf{q})}{v_T^2 q^2 - \omega^2 - \Sigma_T(\mathbf{q},\omega)}\label{eq:green_vec}
\end{align}
with (bare) longitudinal and transverse speed of sound $v_L = \sqrt{(A_0+2B_0)/\rho_0}$ and $v_T = \sqrt{B_0/\rho_0}$ and self-energies $\Sigma_{L,T}$ associated with longitudinal and transverse waves, respectively.

Following the similar argument in the 1-dimensional case, up to the lowest non-zero order in the Born approximation and long-wavelength limit, the renormalized dispersion relations and damping rates are (see details in Sec. \ref{App:B} in Appendix)
\begin{subequations}
\label{eq:vector_results}
\begin{align}
  \Omega_{L,T}^2(q) &= \left[
    1 - \sum_{a,b\in\{A,B\}} C^{(L,T)}_{ab}\,\sigma_{ab}^2\,
    \
  \right]\omega_{0,L,T}^2(q)
  \\
  \Gamma_{L,T}(q) &\propto q^{d+1}
  \sum_{a,b\in\{A,B\}} C^{(L,T)}_{ab}\,\sigma_{ab}^2
\end{align}
\end{subequations}
where $\omega_{0,L,T}(q)=v_{L,T}q$ are homogeneous dispersions. Same as in the 1-dimensional case, both longitudinal and transverse sound velocities are reduced by elastic heterogeneity, while the damping rate exhibits Rayleigh-like scaling, $\Gamma\propto q^{d+1}$. The coefficients $C^{(L,T)}_{ab}$ depend only on $A_0$, $B_0$, and the spatial dimension $d$. Note that the transverse dispersion and damping depend solely on the shear modulus $B_0$, whereas the longitudinal ones involve both $B_0$ and $A_0$.

\subsection{Born-level approximation of elastic wave in density-nonuniform medium}
The same formalism also applies to density disorder \cite{John1983}. For simplicity, we only consider the longitudinal waves traveling in $d$-dimensional media consisting of atoms with a spatially varying density $\rho(\mathbf{r})$ and constant ``spring stiffness" $K_0$. The elastic equation is
\begin{equation}
    \rho(\mathbf{r})\frac{\partial^2 u}{\partial t^2}=K_0\nabla^2u,
\end{equation}
with $\nabla^2=\sum_{i=1}^d\partial^2/\partial r_i^2$ the Laplacian. The random density fluctuates around a mean value $\rho_0$, 
\begin{equation}
    \rho(\mathbf{r})=\rho_0\left(1+\delta\rho(\mathbf{r})\right)
\end{equation}
with $\langle \delta\rho(\mathbf{r})\rangle=0$ and is characterized by spatial correlations, $\langle \delta\rho(\mathbf{r}_0)\delta\rho(\mathbf{r}_0+\mathbf{r})\rangle=\sigma^2\delta^{(d)}(\mathbf{r})(2\pi/q_D)^d$. Likewise, the lowest order Born series approximation yields the equation of average Green's function in terms of self energy (see detailed derivations in Sec. \ref{App:C} in Appendix)
\begin{align}
    \langle G(\mathbf{q},\omega)\rangle&=\frac{1}{-\omega^2+v_L^2q^2-\Sigma(\mathbf{q},\omega)},\\
    \Sigma(\mathbf{q},\omega)&= \frac{\omega^4\sigma^2}{V\rho_0^2}\int_V d^dp G_0(\mathbf{p},\omega),
\end{align}
where $G_0(\mathbf{p},\omega)=G_0(p,\omega)$ is defined in Eq. \eqref{eq:bareGreen}, $v_L=\sqrt{K_0/\rho_0}$ is the longitudinal speed of sound for free wave and $V=q_D^d$ is the total volume of the reciprocal space. In the small $q$ limit, $\omega\propto q\rightarrow0$, the real and imaginary part of the self-energy reads
    \begin{align}
        \text{Re}[\Sigma]\propto q^4\sigma^2, \quad        \text{Im}[\Sigma]\propto q^{d+2}\sigma^2.
        \label{eq:massdisorder}
    \end{align}
In contrast to the heterogeneous elastic case, Rayleigh-type damping also exists in a density-inhomogeneous medium. In other words, Rayleigh damping receives two contributions: heterogeneous elasticity or density. They can lead to the same magnitude. On the other hand, the renormalization in sound propagation speed is of order $\sim\mathcal{O}(q^2)$. 

We further remark that, in the continuum theory, there are only two sources of randomness: density and elastic moduli. Other types of disorder, e.g., positions, are due to the discreteness of the practical lattice pattern and should not contribute to genuine attenuation of the elastic wave - see an example below. 

\section{Numerical methods}
\label{end:mat}
\subsection{MD simulation}
Systems studied in this paper are linear chains and planar (triangular, square) lattices, where particles are connected by springs and are placed such that all particles are in mechanical equilibrium and the whole system is free of internal stresses. The temperature is held at zero, and periodic boundary conditions are imposed.

To measure the phonon transport, we employ the numerical simulation method used in Ref. \cite{Gelin2016,Mizuno2018}, in which the vibrational dynamics around the equilibrium configuration
(inherent structure) are analyzed. For particle $I$, its equilibrium position and displacement are denoted as $\mathbf{r}_I$ and $\mathbf{u}_I$, respectively. At $t=0$, a phonon is excited by perturbing the velocity $\dot{\mathbf{u}}_I$ of particle $I$ in the way that
\begin{equation}
    \dot{\mathbf{u}}_I(t=0)=\mathbf{a}_{L,T}\sin(\mathbf{q}\cdot\mathbf{r}_I+\phi),
\end{equation}
where $L,T$ refer to longitudinal and transverse waves with respect to the wavevector $\mathbf{q}$, $\mathbf{a}=\mathbf{q}/q$ is a unit vector along the $\mathbf{q}$-direction and $\phi=0$ or $\pi/2$ is a constant. Next, we solve the linearized equation of motion for velocities,
\begin{equation}
    \ddot{\mathbf{u}}_I=\sum_{J=1}^N D\mathbf{u}_J,
\end{equation}
where $D$ is the dynamical matrix (see below). Over the time history $\mathbf{u}_I(t)$,  the
normalized velocity-velocity correlation function is calculated as
\begin{equation}
    C(t)\equiv\frac{\sum_{I=1}^N\dot{\mathbf{u}}_I(t)\cdot\dot{\mathbf{u}}_I(0)}{\sum_{I=1}^N\dot{\mathbf{u}}_I(0)\cdot\dot{\mathbf{u}}_I(0)},
\end{equation}
representing the propagation and attenuation behaviors of initially excited phonons. Averaging over the ensemble of different realizations, the final value of $C(t)$ for a given $q$ and polarization is fitted by
\begin{equation}
    C(t)=\cos[\Omega(q)t]e^{-\Gamma(q)t/2},
    \label{eq:corre}
\end{equation}
where the dispersion $\Omega(q)$ and attenuation coefficient $\Gamma(q)$ are extracted. 

\begin{figure}[tbp]
\centering
\includegraphics[width=0.8\textwidth]{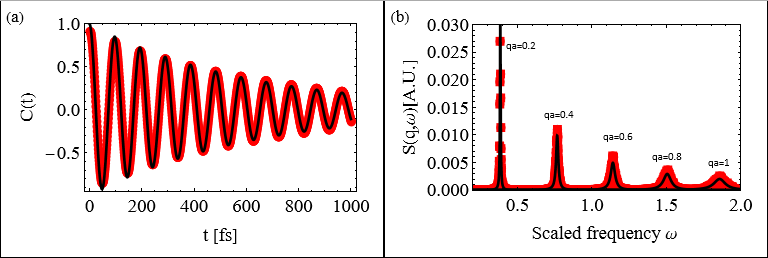}
\caption{Sound attenuation in 1D chains with random springs following an identical independent uniform distribution with mean value $k_0$. Masses are the same $m$ and lattice spacing is $a$. Panel (a) shows the (dimensionless) velocity autocorrelation function $C(t)$ at standard deviation $\sigma=0.25$ and $qa=1$. Red lines are MD data, fitted by $C(t)$ in Eq. \eqref{eq:corre} in the black solid curve.  Panel (b) presents the numerical calculation of the dynamical structure factor (in arbitrary units) for several values of $qa$, cf. Eq. \eqref{eq:dsfnormal}, for the same chain settings in panel (a), plotted against the frequency scaled by a factor $\sqrt{k_0/m}$. Data is represented by red scattered points, while black solid curves refer to the analytical (DHO) fit, cf. Eq. \eqref{eq:DHO}. In both panels, results are obtained after averaging over 40 realizations.}
  \label{fig:1d_md_dsf}
\end{figure}

In Fig. \ref{fig:1d_md_dsf}(a), we present an example of the average velocity correlation of a sample of disordered chains.

\subsection{Dynamical structure factor}
In the classical limit, the Green functions of elastic waves are related to the dynamical structure factor (DSF) by the fluctuation-dissipation theorem \cite{Kubo1957,Schirmacher2015,Schirmacher2015b},
\begin{equation}
     S(\mathbf{q},\omega) =\frac{k_BTq^2}{\pi\omega}\text{Im}\,  \left[G(\mathbf{q},\omega) \right].
\end{equation}
In the one-phonon process, the DSF is a sum over normal modes, taking the form
\begin{equation}
    S(\mathbf{q},\omega)=\frac{\omega k_BT}{N}\sum_s\left|\sum_I\frac{\mathbf{q}\cdot e_I^s(\mathbf{q})}{\omega_s(\mathbf{q})\sqrt{m_I}}\mathbf{e}^{i\mathbf{q}\cdot\mathbf{R}_I}\right|^2\delta(\omega^2-\omega_s^2(\mathbf{q})),
    \label{eq:dsfnormal}
\end{equation}
where $N$ is the total number of particles within the unit cell, $s$ labels normal modes, $\mathbf{e}_I^s(\mathbf{q})$ and $\omega_s(\mathbf{q)}$ are the eigenvector and eigenvalues of the dynamical matrix elements calculated at the equilibrium configuration, i.e.,
\begin{equation}
    D_{IJ}^{\alpha\beta}=\frac{1}{\sqrt{m_Im_J}}\frac{\partial^2\mathcal{U}}{\partial r_I^\alpha\partial r_J^\beta},
\end{equation}
where $\alpha,\beta$ are Cartesian components, $m_I$ is the mass of the $I$th particle, $
\mathcal{U}$ is the potential. For long-wavelength acoustic waves propagating in isotropic random media, the DSF can be approximated by the form of a damped harmonic oscillator (DHO) \cite{Schirmacher2015b}
\begin{equation}
    \langle S(q,\omega)\rangle_{L,T}\propto\frac{k_BTq^2}{\omega\pi}\frac{\omega\Gamma_{L,T}(q)}{(\Omega^2_{L,T}(q)-\omega^2)^2+\omega^2\Gamma_{L,T}(q)^2},
    \label{eq:DHO}
\end{equation}
where $\langle...\rangle$ denotes an ensemble average.
In Fig. \ref{fig:1d_md_dsf}(b), we show a fitting of DSF for a 1D disordered chain with Eq. \eqref{eq:DHO}. We have verified that both methods, MD simulation and normal mode analysis, produce the same results for elastic and mass-disordered harmonic networks \footnote{An exception is for the positional disorder. See Fig. \ref{fig:1d_md_dsf}.}.

\section{Numerical results and discussion}
In this section, we perform numerical calculations on model systems to verify the theoretical calculations in the previous section. There are two ways to extract the information on wave propagation: Molecular Dynamics (MD) simulation and dynamical structure factor based on the eigenmodes (of the dynamical matrix). They are summarized in \textit{Numerical Methods}. 

\subsection{Disordered linear chains}
\begin{figure}[tbp]
\centering
\includegraphics[width=0.8\textwidth]{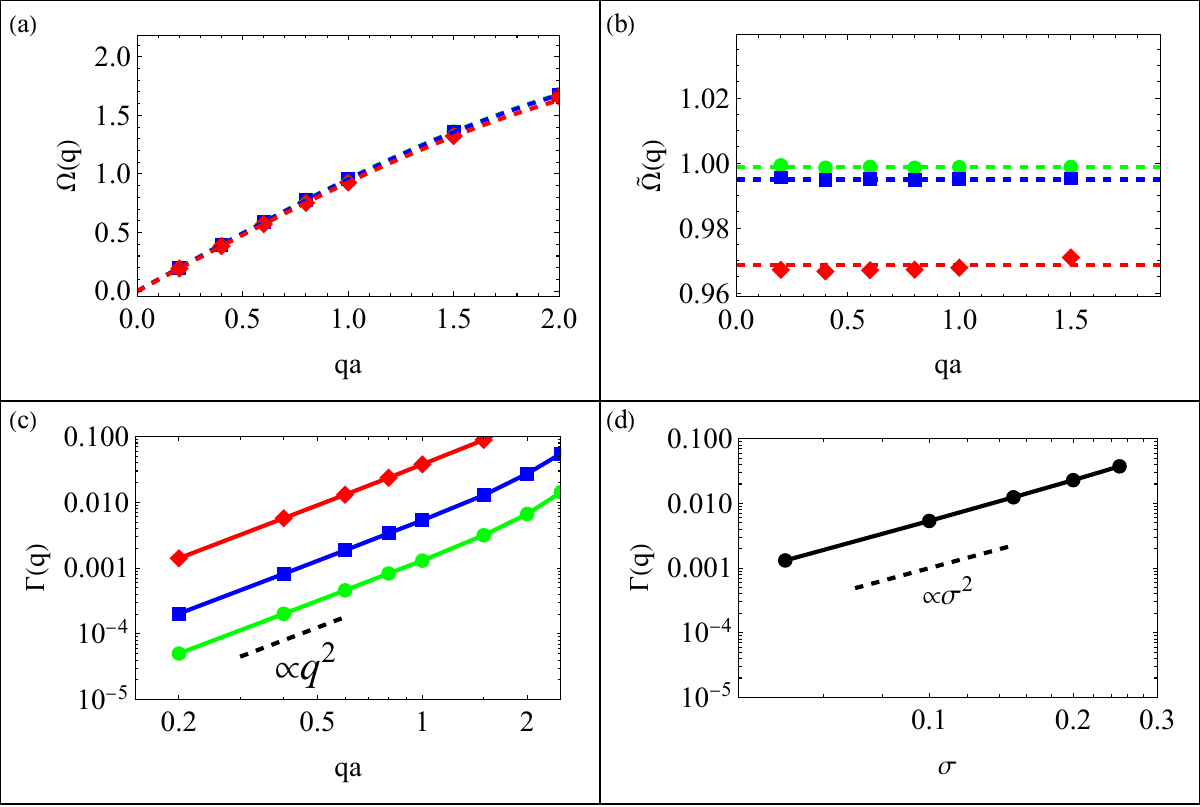}
\caption{The propagation of the sound wave in relaxed 1D chains consisting of $20000$ lattice points of the same mass $m$ connected by independent uniform random springs with the mean spring constant $k_0$ and standard deviation $\sigma$. Panel (a) displays the dispersion $\Omega$ versus $qa$, where $a$ is the lattice spacing. Dotted curves are the result of the Born prediction, e.g. $\Omega\simeq (1-\sigma^2/2)\,\omega_0(q)$ with $\omega_0(q)=2\sin(qa/2)$. Panel (b) shows the ratio of $\Omega$ to the dispersion in a perfect chain, $\tilde{\Omega}=\Omega/\omega_0$ vs. $qa$. Panel (c) depicts the damping rate $\Gamma$ vs. $qa$ and panel (d) indicates $\Gamma$ on the disorder strength at a fixed $qa=1$, respectively. In all panels, scatter points are obtained from the MD simulation after averaging over 40 realizations, where green, blue and red points/lines correspond to $\sigma=0.05, 0.1$ and $0.25$, respectively. In panel (a), green and blue dots/lines overlap. The unit of $\Omega$ and $\Gamma$ is $\sqrt{k_0/m}$.}
  \label{fig:1d_md}
\end{figure}

Figure~\ref{fig:1d_md} collects the MD validation of Eq.~\eqref{eq:scalar_1d_result} for a (relaxed) monatomic chain consisting of equally spaced (spacing $a$), same masses $m$, where the nearest neighbors are connected by uniformly random springs. Consistent with Eq. \eqref{eq:scalar_1d_result}, in the long-wavelength regime, $qa\lesssim 1$, $\Omega(q)\sim(1-\sigma^2/2)vq$, cf. Figs.~\ref{fig:1d_md}(a-b), and Rayleigh damping: $\Gamma(q)\propto q^2$, is observed with $\Gamma\propto\sigma^2$ at fixed $q$, cf. Figs.~\ref{fig:1d_md}(c-d). Importantly, these results depend only on the second-order statistics of $\delta K(x)$ in the weak-disorder limit and therefore hold for arbitrary disorder distributions with the same variance. We have checked that repeating the simulations with normal or binary (two-point) disorder yields identical results (not shown). We have also verified that equivalent results are obtained through the dynamical structure factor. 

\subsection{Disordered planar lattices}
We then test our predictions on an ensemble of 2-dimensional triangular lattices with the same mass $m$ but independent random springs (bonds). The coarse-grained Lam\'e moduli $A$ and $B$ inherit fluctuations in the random spring constants, and the long-wavelength Born predictions are
$\Omega_{L,T}\simeq (1-\sigma^2/3)\,\omega_{0,L,T}(q)$, with $\sigma^2$ the variance of spring constant fluctuation and $\omega_{0,L,T}(q)$ dispersions of the corresponding perfect triangular crystal: For the wave propagating along the bond direction, the acoustic modes split into longitudinal and transverse branches with (in units $\sqrt{k_0/m}, k_0$ the spring constant), 
\begin{subequations}
    \begin{align}
    \omega_{0,L}^2(q) &= 5 - 4\cos^2\left(\frac{qa}{2}\right)
  - \cos\left(\frac{qa}{2}\right),\\
  \omega_{0,T}^2(q) &= 3 - 3\cos\left(\frac{qa}{2}\right),
  \label{eq:tri-disp}
\end{align}
\end{subequations}
where $a$ is the lattice constant. Attenuation coefficient is again Rayleigh-like, $\Gamma_{L,T}\propto \sigma^2 q^3$.

\begin{figure}[tbp]
\centering
\includegraphics[width=0.8\textwidth]{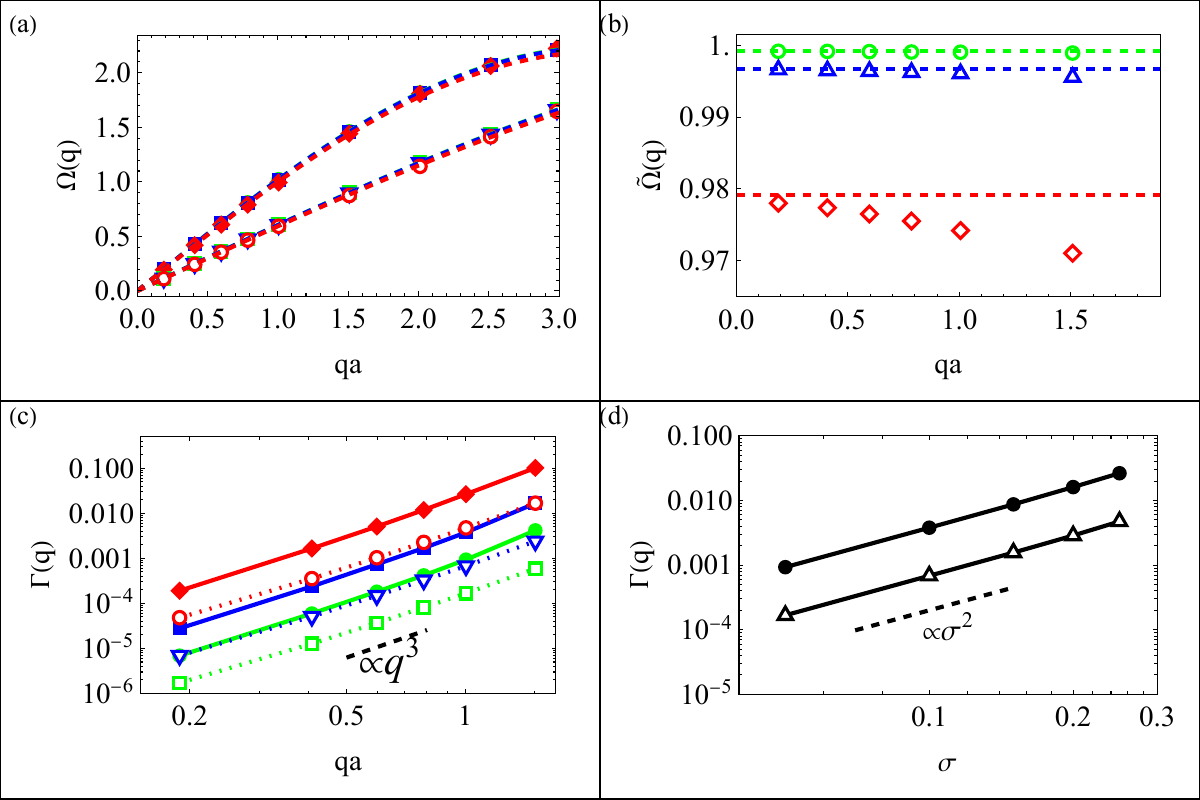}
\caption{The propagation of sound waves in relaxed triangular spring networks consisting of $40000$ lattice points of the same mass $m$ connected by independent uniform random springs with the mean spring constant $k_0$ and standard deviation $\sigma$. Panel (a) displays the dispersion $\Omega_{L,T}$ versus $qa$ where $a$ is the lattice spacing. The dotted curve is the result of the Born prediction, cf. Eq. \eqref{eq:vector_results}. Panel (b) shows the ratio of $\Omega_T$ to the (transverse) dispersion in a perfect lattice, $\tilde{\Omega}_T=\Omega_T/\omega_{0,T}$ vs. $qa$. Panel (c) shows the damping rate $\Gamma_{L,T}$ vs. $qa$. Panel (d) indicates $\Gamma_{L,T}$ on the disorder strength at a fixed $qa=1$. In all panels, scatter points are obtained from the MD simulation after averaging over 40 realizations, where phonons are excited with wavevector along the bond direction. Filled dots are longitudinal waves, while blank points refer to transverse waves, where green, blue and red correspond to $\sigma=0.05, 0.1$ and $0.25$, respectively. In panel (a), green and blue dots/lines overlap. The unit of $\Omega$ and $\Gamma$ is $\sqrt{k_0/m}$. }
  \label{fig:2d_md}
\end{figure}

Figure~\ref{fig:2d_md} compares Born predictions Eq. \eqref{eq:vector_results} with the simulation data: In the long-wavelength regime the MD-extracted $\Omega_{L,T}(q)$ and $\Gamma_{L,T}(q)$ closely follow the analytic predictions, showing that both longitudinal and transverse branches soften as $\sigma$ increases and that the attenuation obeys the Rayleigh law $\Gamma\propto \sigma^2 q^3$ expected for independent weak disorder. As $qa$ grows, the long-wavelength and isotropy assumptions break down, the lattice anisotropy becomes resolvable, and discrepancies with the continuum theory naturally appear. We have also verified that the same $1-\sigma^2/3$ sound-speed renormalization and Rayleigh damping also describe a more anisotropic 2D square lattice with random nearest- and next-nearest-neighbor springs (see results in Sec. \ref{App:D} in Appendix), indicating that the leading Born correction is largely insensitive to the underlying lattice symmetry in the low $q$ regime.

\subsection{Disordered crystals with random masses}
\begin{figure}[tbp]
\centering
\includegraphics[width=0.8\textwidth]{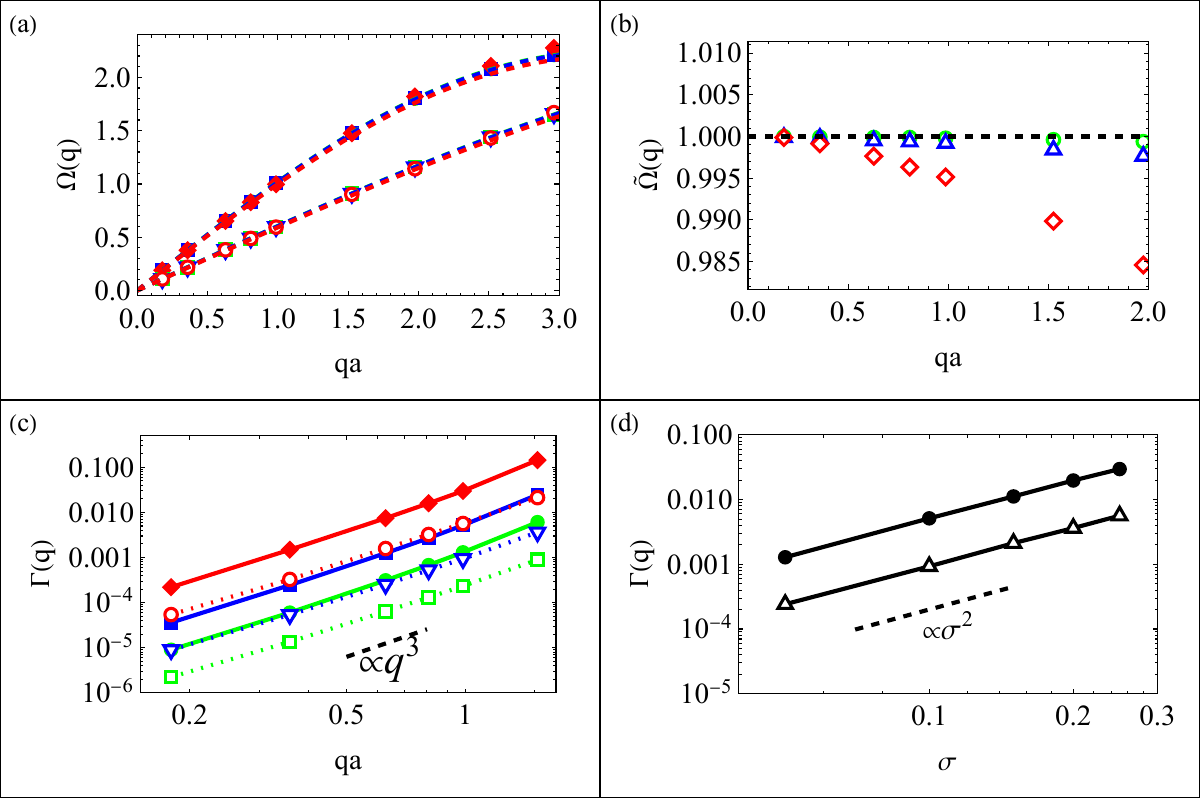}
\caption{The propagation of sound waves in relaxed triangular spring bonds consisting of $40000$ lattice points with different masses fluctuating uniformly around a mean mass $m_0$ with standard deviation $\sigma$. Nearest neighbors are connected by the same springs. Panel (a) displays the dispersion $\Omega_{L,T}$ versus $qa$ where $a$ is the lattice spacing. The dotted curve is the result of the Born prediction, cf. Eq. \eqref{eq:vector_results}. Panel (b) shows the ratio of $\Omega$ to the (transverse) dispersion in a perfect lattice, $\tilde{\Omega}=\Omega/\omega_0$ vs. $qa$. Panel (c) shows the damping rate $\Gamma_{L,T}$ vs. $qa$. Panel (d) indicates $\Gamma_{L,T}$ on the disorder strength (standard deviation) at a fixed $qa=1$. In all panels, scatter points are obtained from the MD simulation after averaging over 40 realizations, where phonons are excited with wavevector along the bond direction. Filled dots are longitudinal waves, while blank points refer to transverse waves, where green, blue and red correspond to $\sigma=0.05, 0.1$ and $0.25$, respectively. In panel (a), green and blue dots/lines overlap. The unit of $\Omega$ and $\Gamma$ is $\sqrt{k_0/m_0}$. }
  \label{fig:2d_md_m}
\end{figure}

Taking the same lattices in (1- and) 2-dimensional systems as in the previous sections, where the springs are identical but the particle masses are drawn independently from uniform distributions with variance $\sigma^2$, the extracted damping rate still shows Rayleigh scaling, $\Gamma(q)\propto \sigma^2 q^{d+1}$ in the long-wavelength regime, whereas the dispersion is essentially unchanged. These can be seen in Fig. \ref{fig:2d_md_m}, which is consistent with predictions in Eq. \eqref{eq:massdisorder}.

\begin{figure}[tbp]
\centering
\includegraphics[width=0.8\textwidth]{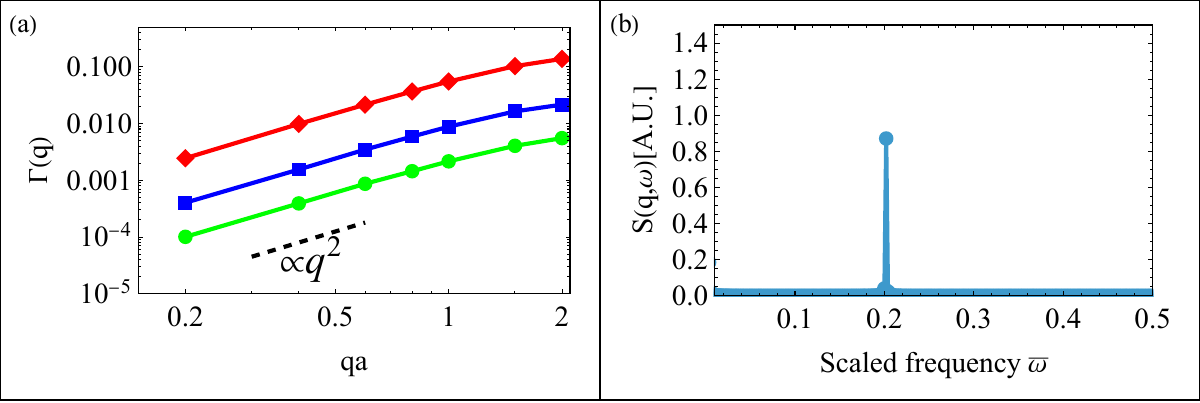}
\caption{The propagation of the sound wave in relaxed 1D chains consisting of $20000$ lattice points of the same mass $m$ connected by the springs with uniformly random natural lengths around the mean spacing $a$. Panel (a) displays the damping rate of $\Gamma$ versus $qa$. Scatter points are obtained from the MD simulation after averaging over 40 realizations, where green, blue and red correspond to the standard deviation $\sigma=0.05, 0.1$ and $0.25$ (in units of $a$), respectively. Panel (b) presents the calculation of the dynamical structure factor for the same (positional) disordered chains with $\sigma=0.25$ (in units of $a$).}
  \label{fig:md_dsf_p}
\end{figure}

\subsection{Positional disorder}
We further simulate realizations of spring chains presenting quenched disorder in equilibrium bond lengths, keeping all masses and spring constants equal. Note that random numbers are also natural lengths of springs since the system should be in equilibrium and relaxed. As is seen in Fig. \ref{fig:md_dsf_p}(a), MD measurements again produce an apparent Rayleigh-like $\Gamma(q)$ of comparable magnitude to the spring- (and mass-) disorder cases.
Because the dynamics (displacements) only concern motions relative to equilibrium positions, however, the dynamical matrix of this positional-disorder chain is the same as that of the ordered chain. Consequently, no damping is expected in the dynamical structure factor and hence in the sound wave, as is shown in Fig. \ref{fig:md_dsf_p}(b). Therefore, in structural-disordered harmonic bonds, the damping inferred from MD, using the usual protocol of exciting a phonon via an initial velocity profile $v_n\propto \sin(qx_n)$ at the positions $x_n$ of the $n$th particle, should be viewed as an artefact of the phonon excitation protocol rather than genuine phonon scattering. A similar conclusion can be made for calculations within nonaffine elasticity theory \cite{Cui2020c}: The random rest lengths generate a non-vanishing affine force field, but the resulting nonaffine correction to the elastic constant exactly cancels the change in the affine contribution, so the net elastic constant remains identical to that of the perfect chain. In other words, attenuation is not necessarily caused by nonaffine dynamics. 

\section{Conclusion}
In this work, tracking the lowest order approximation of the Born series, we derived coupled equations between Green's functions and self-energies, obtaining closed expressions that relate reduced dispersions and Rayleigh-type attenuation directly to the variance of the elastic disorder in the long-wavelength limit. MD simulations and exact diagonalization of disordered 1D chains and 2D lattices quantitatively confirm these predictions.

Complementary simulations for 1-dimensional chains with mass disorder show clear Rayleigh-like damping while the sound speed remains essentially unchanged, demonstrating that Rayleigh scaling is not specific to modulus fluctuations. For harmonic spring networks with positional disorder, the dynamical matrix (and dynamical structure factor) is identical to that of the perfect crystal, implying that the apparent damping seen in MD might be a consequence of the phonon excitation protocol, rather than as evidence of genuine scattering. 

While we focus on spatially uncorrelated quenched disordered solids, our approach can be easily extended to spatially correlated situations that are more closely related to real materials \cite{Gelin2016,Lindsay2023}. Additionally, the practical measurement of amorphous materials spans a broad range of wavelengths. Extending to large disorder at a smaller scale (intermediate wavenumber) might be achieved by, for example, solving complete self-consistent equations. Comparing theoretical predictions with phonon transport in real amorphous materials would be a desirable future direction. 


\begin{acknowledgments}
Insightful discussions with Matteo Baggioli are greatly acknowledged. B.C. thanks Maria for being around when writing the paper. B.C. acknowledges the financial support of the National Natural Science Foundation of China (No. 12404232), start-up funding from the Chinese University of Hong Kong, Shenzhen (No. UDF01003468) and the Shenzhen city “Pengcheng Peacock” Talent Program.
\end{acknowledgments}

\appendix

\section{Heterogeneous elastic theory in 1 dimension}
\label{App:A}
In this section, we derive the form for the self-energy of sound wave propagation in a heterogeneous elastic continuum in 1-dimensional space, starting from the equation, cf. Eq. (1) in the main text,
\begin{equation}
  \rho_0\frac{\partial^2u}{\partial t^2} = \frac{\partial}{\partial x}\left[K(x)\frac{\partial u}{\partial x}\right].
  \label{eq:1d_wave}
\end{equation}
The Green function $G(x,x_0,t,t_0)$ associated with Eq. \eqref{eq:1d_wave} satisfies 
\begin{equation}
    \left[\rho_0\frac{\partial^2G}{\partial t^2}+\frac{\partial}{\partial x}\left(K(x)\frac{\partial G}{\partial x}\right)\right]=\delta(t-t_0)\delta(x-x_0).
\end{equation}
For the displacement field $u(x,t)$, its Fourier transform is
\begin{equation}
    u(q,\omega)=\int u(x,t)e^{-i\omega t-iqx}dxdt.
\end{equation}
In Fourier space, Eq. \eqref{eq:1d_wave} becomes
\begin{equation}
    (-\omega^2+v_0^2q^2)G(q,\omega)-\frac{1}{\rho_0}\int_0^{q_D}\frac{dp}{2\pi}qp\delta K(q-p)G(p,\omega)=1,
    \label{eq:FTGreen}
\end{equation}
where $\delta K(q)$ denotes the Fourier transform of $\delta K(x)$, $q_D$ is the Debye wavenumber, $v_0 = \sqrt{K_0/\rho_0}$ is the speed of sound in the homogeneous medium. Denoting the phonon Green's function of the homogeneous system,
\begin{equation}
  G_{0}(q,\omega) = \frac{1}{-\omega^2 + v_0^2 q^2 + i0^+},
  \label{eq:bareGreen}
\end{equation}
and multiplying both sides of  Eq. \eqref{eq:FTGreen} by $G_{0}(q,\omega)$ yield
\begin{align}
    G(q,\omega)&=G_0(q,\omega)+G_0(q,\omega)\int_0^{q_D}\frac{dp}{2\pi}V_{qp}G(p,\omega),\\
    V_{qp}&\equiv\frac{qp}{\rho_0}\delta K(q-p).
\end{align}
The introduction of elastic heterogeneity breaks continuous translational symmetry, causing the scattering of plane waves, which is described by the scattering potential $V_{qp}$, where $q$ and $p$ denote incoming and outgoing wavevectors, i.e., the elastic heterogeneity induces scattering with momentum transfer $k=q-p$ and amplitude determined by $\delta K(k)$. For homogeneous media ($\delta K=0$), only $k=0$ exists, i.e., no scattering.

The full Green's function can be iterated to yield the Born series
\begin{align}
  G(q,\omega) &= G_{0}(q,\omega)
  + G_{0}(q,\omega)\int_0^{q_D}\frac{dp}{2\pi} \,V_{qp}\,G_{0}(p,\omega)
  \nonumber\\
  &+ G_{0}(q,\omega)\int_0^{q_D}\int_0^{q_D}\frac{dp\,dk}{(2\pi)^2}
  \,V_{qp}\,G_{0}(p,\omega)\,V_{pk}\,G_{0}(k,\omega)
    \nonumber\\
  &+ \cdots
\end{align}
Using the operator notation, it can be noted as $G=G_0+G_0VG_0+G_0VG_0VG_0+...$, which, after taking the ensemble average over disorder and keeping only the 2nd order of $\delta K$, becomes
\begin{equation}
    \langle G\rangle=G_0+G_0\langle VG_0V\rangle G_0.
    \label{eq:G2ndorder}
\end{equation}
Note that this lowest-order expansion is in general valid for small fluctuations. On the other hand, the full Green function $G$ might be expressed upon the introduction of self-energy,
\begin{equation}
    \langle G(q,\omega)\rangle=\left[G_0^{-1}(q,\omega)-\Sigma(q,\omega)\right]^{-1}.
    \label{eq:fullGself}
\end{equation}
Comparing Eqs. (\ref{eq:G2ndorder}-\ref{eq:fullGself}), one can approximate the self-energy by noticing that
\begin{align}
    &\int_0^{q_D}\int_0^{q_D}\frac{dpdk}{(2\pi)^2}\langle V_{qp}G_0(p,\omega)V_{pk}\rangle G_0(k,\omega)\notag\\
    =&\frac{1}{2\pi\rho_0^2}\int_0^{q_D}q^2p^2\eta(q-p)G_0(p,\omega)G_0(q,\omega),
\end{align}
where 
\begin{equation}
    \eta(q)=\int e^{-iqx}\langle\delta K(x_0)\delta K(x_0+x)\rangle dx,
\end{equation}
is the Fourier transform of the elastic correlation. When the elastic fluctuation is spatially uncorrelated, i.e. $\langle\delta K(x_0)\delta K(x_0+x)\rangle=\sigma^2\delta(x)2\pi/q_D$, and up to the variance, the Born approximation of the self-energy reads 
\begin{equation}
    \Sigma_{\text{Born}}(q,\omega)=\frac{\sigma^2q^2}{q_D\rho_0^2}\int_0^{q_D}p^2G_0(p,\omega)dp,
    \label{eq:sigmawithG0}
\end{equation}
which is Eq. (4) in the main text. Using the Plemelj formula, the free Green function may be expressed as
\begin{equation}
    G_0(p,\omega)=\mathcal{P}\frac{1}{v_0^2p^2-\omega^2}-i\pi\delta(v_0^2p^2-\omega^2),
    \label{eq:Plemelj}
\end{equation}
where $\mathcal{P}$ means the principal part. Substituting $G_0$ into Eq. \eqref{eq:sigmawithG0} and take the long-wavelength limit $\omega\sim v_0q\rightarrow0$, the real and imaginary parts of the self-energy reads
\begin{align}
    \text{Re}[\Sigma_{\text{Born}}(q,\omega)]&=\frac{q^2\sigma^2}{\rho_0^2v_0^2}+\mathcal{O}(q^4),\\
    \text{Im}[\Sigma_{\text{Born}}(q,\omega)]&=-\frac{\pi\sigma^2q^3}{2q_D\rho_0^2},
\end{align}
leading to Eq. (7) in the main text.

\section{Heterogeneous elastic theory in higher dimensions}
\label{App:B}
For spatial dimension $d\ge 2$, the displacement field is vectorial. The $\alpha$th component of the displacement field in an isotropic elastic medium obeys
\begin{equation}
  \rho_0\frac{\partial^2 u_\alpha}{\partial t^2} =\sum_{\beta\gamma\chi} \frac{\partial}{\partial r_\beta} 
  \left[C_{\alpha\beta\gamma\chi}(\mathbf{r})\frac{\partial u_\chi}{\partial r_{\gamma}}\right] 
  \label{eq:vec_wave}
\end{equation}
with
\begin{equation}
  C_{\alpha\beta\gamma\chi}(\mathbf{r}) = A(\mathbf{r})\,\delta_{\alpha\beta}\delta_{\gamma\chi}
  + B(\mathbf{r})\left(\delta_{\alpha\gamma}\delta_{\beta\chi}
  + \delta_{\alpha\chi}\delta_{\beta\gamma}\right)
  \label{eq:lame}
\end{equation}
where $A(\mathbf{r}) = A_0\bigl[1+\delta A(\mathbf{r})\bigr]$ is Lam\'e's first parameter and $B(\mathbf{r}) = B_0\bigl[1+\delta B(\mathbf{r})\bigr]$ is the shear modulus, whose random parts fluctuate in the way that $\langle \delta a(\mathbf{r})\rangle=0, \langle \delta a(\mathbf{r}_0)\delta b(\mathbf{r}_0+\mathbf{r})\rangle=\sigma^2_{ab}\delta^{(d)}(\mathbf{r})(2\pi/q_D)^d, a=A,B$.  Following the same treatment as in the 1D case, we attempt to look for Green's function $G_{\alpha\beta}(\mathbf{r},t)$ satisfying
\begin{equation}
    \sum_{\beta}\left[\hat{L}_{\alpha\beta}^{(0)}+\hat{V}_{\alpha\beta}\right]G_{\beta\gamma}(\mathbf{r},t)=\delta_{\alpha\gamma}\delta^{(d)}(\mathbf{r})\delta(t),
    \label{eq:isotensor}
\end{equation}
with
\begin{align}
    \hat{L}_{\alpha\beta}^{(0)}&=\left[\rho_0\frac{\partial^2}{\partial t^2}\delta_{\alpha\beta}-(A_0+B_0)\frac{\partial^2}{\partial r_\alpha\partial r_\beta}-B_0\delta_{\alpha\beta}\nabla^2\right],\\
    \hat{V}_{\alpha\beta}&=-\left[\frac{\partial \delta A(\mathbf{r})}{\partial r_\alpha}\frac{\partial}{\partial r_\beta}+\frac{\partial \delta B(\mathbf{r})}{\partial r_\beta}\frac{\partial }{\partial r_\alpha}+\delta_{\alpha\beta}\sum_{\gamma}\frac{\partial\delta B(\mathbf{r})}{\partial r_\gamma}\frac{\partial}{\partial r_\gamma}\right].
\end{align}
Making the Fourier transform, e.g.,
\begin{equation}
    G_{\alpha\beta}(\mathbf{q},\omega)=\int e^{-i\mathbf{q}\cdot\mathbf{r}-i\omega t}G_{\alpha\beta}(\mathbf{r},t)d^drdt,
\end{equation}
Eq. \eqref{eq:isotensor} becomes
\begin{equation}
    \sum_{\beta}\left[L_{\alpha\beta}^{(0)}(\mathbf{q},\omega)G_{\beta\gamma}(\mathbf{q},\omega)+\int\frac{d^dp}{(2\pi)^d}V_{\alpha\beta}(\mathbf{q},\mathbf{p})G_{\beta\gamma}(\mathbf{p},\omega)\right]=\delta_{\alpha\gamma},
    \label{eq:FTisotensor}
\end{equation}
where the free operator $\hat{L}_{\alpha\beta}^{(0)}$ may be written as
\begin{equation}
    L_{\alpha\beta}^{(0)}(\mathbf{q},\omega)=(B_0q^2-\rho_0\omega^2)\delta_{\alpha\beta}+(A_0+B_0)q_\alpha q_\beta,
\end{equation}
which, in terms of projectors $P^{L}_{\alpha\beta}(\mathbf{q})\equiv q_\alpha q_\beta/q^2$ and
$P^{T}_{\alpha\beta}(\mathbf{q})\equiv \delta_{\alpha\beta} - q_\alpha q_\beta/q^2$, is
\begin{equation}
    L_{\alpha\beta}^{(0)}(\mathbf{{q}},\omega)=((A_0+2B_0)q^2-\rho_0\omega^2)P_{\alpha\beta}^L+(B_0q^2-\rho_0\omega^2)P_{\alpha\beta}^T.
\end{equation}
Similarly, the coupling term is
\begin{equation}
    V_{\alpha\beta}(\mathbf{q},\mathbf{p})=-\left[q_\alpha p_\beta\delta A(\mathbf{q}-\mathbf{p})+(p_\alpha q_\beta+(\mathbf{q}\cdot\mathbf{p})\delta_{\alpha\beta})\delta B(\mathbf{q}-\mathbf{p})\right].
    \label{eq:Vab}
\end{equation}
Multiplying Eq. \eqref{eq:FTisotensor} by $\sum_\alpha G^{(0)}_{\alpha\kappa}$, the Green function corresponding to $\hat{L}^{(0)}_{\alpha\kappa}$, yields the Dyson equation
\begin{equation}
    G_{\alpha\beta}(\mathbf{q},\omega)=G_{\alpha\beta}^{(0)}(\mathbf{q},\omega)-\sum_{\alpha\beta}G_{\alpha\beta}^{(0)}(\mathbf{q},\omega)\int\frac{d^dp}{(2\pi)^d}V_{\alpha\beta}(\mathbf{q},\mathbf{p})G_{\alpha\beta}(\mathbf{p},\omega).
\end{equation}
Iterating the Dyson equation and taking into account the correlation of fluctuations of elastic parameters, the average Green function may be approximated as
\begin{equation}
    \langle G_{\alpha\beta}(\mathbf{q},\omega)\rangle\approx G_{\alpha\beta}^{(0)}(\mathbf{q},\omega)+\sum_{\gamma\kappa\mu\nu}G_{\alpha\gamma}^{(0)}(\mathbf{q},\omega)\int\frac{d^dp}{(2\pi)^d}\int\frac{d^dk}{(2\pi)^d} \langle V_{\gamma\kappa}(\mathbf{q},p)V_{\mu\nu}(\mathbf{p},\mathbf{k}\rangle)G_{\kappa\mu}^{(0)}(\mathbf{p},\omega)G_{\nu\beta}^{(0)}(\mathbf{k},\omega).
\end{equation}
Substituting $V_{\alpha\beta}(\mathbf{q},\mathbf{p})$ in Eq. \eqref{eq:Vab}, the 2nd order term is
\begin{equation}
    \frac{1}{q_D^d}\sum_{\gamma\kappa\mu\nu}G_{\alpha\gamma}^{(0)}(\mathbf{q},\omega)\int\frac{d^dp}{(2\pi)^d} W_{\gamma\kappa\mu\nu}(\mathbf{p},\mathbf{q})G_{\kappa\mu}^{(0)}(\mathbf{p},\omega)G_{\nu\beta}^{(0)}(\mathbf{q},\omega)
    \label{eq:selfvector}
\end{equation}
where 
\begin{align}
    W_{\gamma\kappa\mu\nu}(\mathbf{q},\mathbf{p})&=\sigma_{AA}^2q_{\gamma}p_{\kappa}q_{\mu}p_{\nu}\notag\\
    &+\sigma_{BB}^2(p_{\gamma}q_{\kappa}+(\mathbf{q}\cdot\mathbf{p})\delta_{\gamma\kappa})(p_{\mu}q_{\nu}+(\mathbf{q}\cdot\mathbf{p})\delta_{\mu\nu})\notag\\
    &+\sigma_{AB}^2q_{\gamma}p_{\kappa}(p_{\mu}q_{\nu}+(\mathbf{q}\cdot\mathbf{p})\delta_{\mu\nu})\notag\\
    &+\sigma_{BA}^2q_{\mu}p_{\nu}(p_{\gamma}q_{\kappa}+(\mathbf{q}\cdot\mathbf{p})\delta_{\gamma\kappa})\notag\\
    &\sim \mathcal{O}(q^2)\mathcal{O}(p^2)
\end{align}
Note that, making use of the identity 
\begin{equation}
    \left(\frac{P_{\alpha\beta}^L}{a}+\frac{P^T_{\alpha\beta}}{b}\right)(aP_{\alpha\beta}^L+bP_{\alpha\beta}^T)=\delta_{\alpha\beta},
\end{equation}
where $a,b$ are scalars, the free Green function may be decomposed as
\begin{align}
    G_{\alpha\beta}^{(0)}(\mathbf{p,\omega})=\frac{1}{\rho_0}\left[\frac{P^{L}_{\alpha\beta}(\mathbf{p})}{v_L^2 p^2 - \omega^2}+\frac{P^{T}_{\alpha\beta}(\mathbf{p})}{v_T^2 p^2 - \omega^2}\right].
\end{align}
with longitudinal and transverse speed of sound $v_L = \sqrt{(A_0+2B_0)/\rho_0}$ and $v_T = \sqrt{B_0/\rho_0}$, respectively.
Making the same argument as in the scalar case, namely using the Plemelj formula for the free Green function, e.g. Eq. \eqref{eq:Plemelj}, substituting back into Eq. \eqref{eq:selfvector} and taking the long-wavelength limit, the full Green function may be written as
\begin{align}
  \langle G_{\alpha\beta}(\mathbf{q},\omega)\rangle
  &= \frac{P^{L}_{\alpha\beta}(\mathbf{q})}{v_L^2 q^2 - \omega^2 - \Sigma_L(\mathbf{q},\omega)}\nonumber\\
  &+ \frac{P^{T}_{\alpha\beta}(\mathbf{q})}{v_T^2 q^2 - \omega^2 - \Sigma_T(\mathbf{q},\omega)},
  \label{eq:green_vec}
\end{align}
where $\Sigma_{L,T}$ are self-energies corresponding to longitudinal and transverse waves, respectively. Similar to the scalar case, the real and imaginary parts of the self-energy are
\begin{align}
    \text{Re}[\Sigma_{L,T}(q,\omega)]&\propto q^2\sigma^2_{ab}+o(q^2),\\
    \text{Im}[\Sigma_{L,T}(q,\omega)]&\propto-\sigma_{ab}^2q^{d+2}.
\end{align}
These are verified in numerical simulations in the main text.

\section{Lowest-order Born approximation of elastic wave in the mass-disordered medium}
\label{App:C}
In this section, we derive the form for the self-energy of scalar sound wave propagation in a continuum with fluctuating density in $d$-dimensional space. Starting from the wave equation, namely Eq. (12) in the main text,
\begin{equation}
  \rho(\mathbf{r})\frac{\partial^2u(\mathbf{r},t)}{\partial t^2} = K_0\nabla^2u(\mathbf{r},t),
  \label{eq:d_wave_density}
\end{equation}
where 
\begin{equation}
    \rho(\mathbf{r})=\rho_0(1+\delta\rho(\mathbf{r})),
\end{equation}
with $\langle\delta\rho(\mathbf{r})\rangle=0$ and $\langle\delta\rho(\mathbf{r}_0)\delta \rho(\mathbf{r}_0+\mathbf{r})\rangle=f(\mathbf{r})$ for some function $f(\mathbf{r)}$, we make the Fourier transform to work in Fourier space, having
\begin{equation}
    (-\omega^2+v_0q^2)G(\mathbf{q},\omega)-\frac{\omega^2}{\rho_0}\int_V\frac{d^dp}{(2\pi)^d}\delta \rho(\mathbf{q}-\mathbf{p})G(\mathbf{p},\omega)=1,
    \label{eq:FTGreenmass}
\end{equation}

\begin{figure}[h]
\centering
\includegraphics[width=0.8\textwidth]{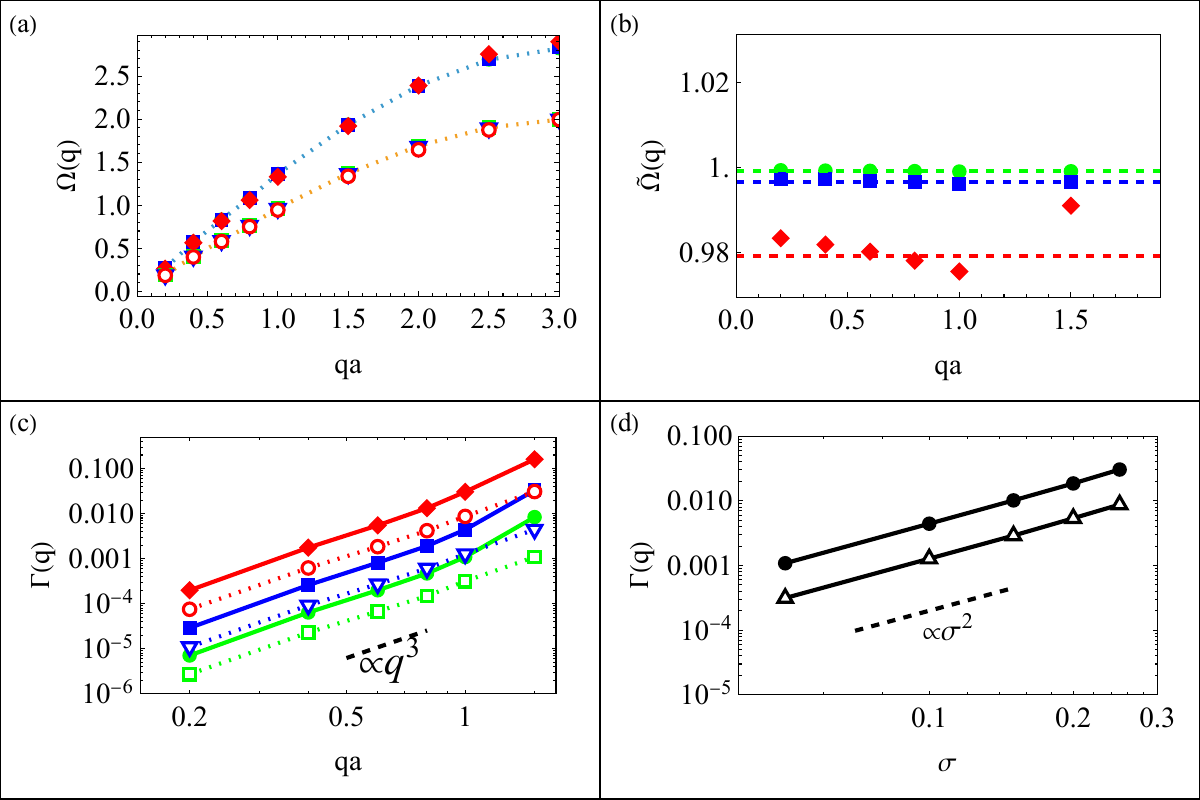}
\caption{The propagation of sound waves in relaxed square spring bonds consisting of $40000$ lattice points of the same mass $m$, where the nearest and next-nearest neighors are connected by independent uniform random springs with the mean spring constant $k_0$ and standard deviation $\sigma$ (in units of $k_0$). Panel (a) displays the dispersion $\Omega_{L,T}$ versus $qa$ where $a$ is the lattice spacing. Dotted curves are the dispersions of a perfect square lattice. Panel (b) shows the ratio of $\Omega$ to the (longitudinal) dispersion in a perfect lattice versus $qa$. Panel (c) shows the damping rate $\Gamma_{L,T}$ vs. $qa$. Panel (d) indicates $\Gamma_{L,T}$ vs. disorder strength (standard deviation) $\sigma$ at a fixed $qa=1$. In all panels, scatter points are obtained from the MD simulation after averaging over 40 realizations, where phonons are excited with wavevector along the shortest bond direction. Filled dots are longitudinal waves, while blank points are transverse waves, where green, blue and red correspond to $\sigma=0.05, 0.1$ and $0.25$, respectively. In panel (a), green and blue dots/lines overlap. The unit of $\Omega$ and $\Gamma$ is $\sqrt{k_0/m}$. }
  \label{fig:S1}
\end{figure}

Following a similar procedure to the heterogeneous elastic case, up to the lowest non-zero order, the full Green function is expressed in terms of the free wave \eqref{eq:bareGreen} and self-energy,
\begin{equation}
    \langle G(\mathbf{q},\omega)\rangle=\left[G_0^{-1}(\mathbf{q},\omega)-\Sigma(\mathbf{q},\omega)\right]^{-1}.
    \label{eq:fullGselfmass}
\end{equation}
with 
\begin{equation}
    \Sigma(\mathbf{q},\omega)\approx\omega^4\int_V\frac{d^dp}{(2\pi)^d}\eta(\mathbf{q}-\mathbf{p})G_0(\mathbf{p},\omega),
\end{equation}
where 
\begin{equation}
    \eta(\mathbf{q})=\int e^{-i\mathbf{q}\cdot\mathbf{r}}\langle\delta \rho(\mathbf{r}_0)\delta \rho(\mathbf{r}_0+\mathbf{r})\rangle d^dr,
\end{equation}
is the Fourier transform of the density correlation. Restricting ourselves to uncorrelated random densities, i.e. $\langle\delta \rho(\mathbf{r}_0)\delta \rho(\mathbf{r}_0+\mathbf{r})\rangle=\sigma^2\delta(\mathbf{r})(2\pi)^d/V$, we have, up to this order,
\begin{equation}
    \Sigma(\mathbf{q},\omega)\approx\frac{\sigma^2\omega^4}{V\rho_0^2}\int_VG_0(\mathbf{p},\omega)d^dp,
\end{equation}
which is Eq. (15) in the main text. In the long-wavelength limit $\omega\sim v_0q\rightarrow0$, the real and imaginary parts of the self-energy become
\begin{align}
    \text{Re}[\Sigma(\mathbf{q},\omega)]&\propto \omega^4\sigma^2,\\
    \text{Im}[\Sigma(\mathbf{q},\omega)]&\propto\omega^{d+2}\sigma^2,
\end{align}
corresponding to Eq. (16) in the main text.

\section{Disordered square lattice}
\label{App:D}
We present MD simulations of sound attenuation in 2D disordered square lattices consisting of the same lattice points, where nearest- and next-nearest neighbours are connected with random springs. The dispersion and attenuation coefficient are shown in Fig. \ref{fig:S1}.

\bibliography{apssamp}

\end{document}